\documentclass[runningheads]{llncs}
\usepackage[T1]{fontenc}
\usepackage{graphicx}
\usepackage{multirow}
\usepackage{hyperref} 
\usepackage{color}
\usepackage{tcolorbox}
\usepackage{booktabs}

\usepackage{orcidlink}

\begin{document}
\title{Testing software for non-discrimination: an updated and extended audit in the Italian car insurance domain}
\titlerunning{An updated and extended audit in the Italian car insurance domain}
%
 \author{Marco Rondina\orcidlink{0009-0008-8819-3623}\inst{1} \and
 Antonio Vetrò\orcidlink{0000-0003-2027-3308}\inst{1} \and Riccardo Coppola\orcidlink{0000-0003-4601-7425}\inst{1} \and Oumaima Regragrui\inst{1} \and
 Alessandro Fabris\orcidlink{0000-0001-6108-9940}\inst{2} \and Gianmaria Silvello\orcidlink{0000-0003-4970-4554}\inst{3} \and Gian Antonio Susto\orcidlink{0000-0001-5739-9639}\inst{3} \and
 Juan Carlos De Martin\orcidlink{0000-0002-7867-1926}\inst{1}}
%
\authorrunning{M. Rondina et al.}
%
 \institute{Politecnico di Torino, Turin, Italy \\
 \email{\{marco.rondina, antonio.vetro, juancarlos.demartin\}@polito.it} \email{oumaima.regragui@studenti.polito.it} \and
 Max Planck Institute for Security and Privacy, Bochum, Germany
 \email{alessandro.fabris@mpi-sp.org}\\
 \and
 Università di Padova, Padua, Italy\\
 \email{\{gianmaria.silvello,gianantonio.susto\}@unipd.it}}
\maketitle              
\begin{abstract}
\textbf{Context.} As software systems become more integrated into society’s infrastructure, the responsibility of software professionals to ensure compliance with various non-functional requirements increases. These requirements include security, safety, privacy, and, increasingly, non-discrimination.

\noindent\textbf{Motivation.} Fairness in pricing algorithms grants equitable access to basic services without discriminating on the basis of protected attributes.

\noindent\textbf{Method.} We replicate a previous empirical study that used black box testing to audit pricing algorithms used by Italian car insurance companies, accessible through a popular online system. With respect to the previous study, we enlarged the number of tests and the number of demographic variables under analysis.

\noindent\textbf{Results.} Our work confirms and extends previous findings, highlighting the problematic permanence of discrimination across time: demographic variables significantly impact pricing to this day, with birthplace remaining the main discriminatory factor against individuals not born in Italian cities. We also found that driver profiles can determine the number of quotes available to the user, denying equal opportunities to all. 

\noindent\textbf{Conclusion}. The study underscores the importance of testing for non-discrimination in software systems that affect people's everyday lives. Performing algorithmic audits over time makes it possible to evaluate the evolution of such algorithms. It also demonstrates the role that empirical software engineering can play in making software systems more accountable. 

\end{abstract}

\section{Introduction and Motivation}
\label{sec:intro}
As the level of integration of software systems into the infrastructure of society rapidly increases, so does the responsibility of software professionals \cite{acmcode2018taskforceACMCodeEthics2018}. They must ensure that the software they deploy meets a wide range of non-functional requirements such as security, safety, privacy, and, increasingly, non-discrimination. Addressing issues of unfairness in different application domains has become a critical and urgent task: recent studies have identified algorithmic discriminations in areas that are crucial for maintaining equal opportunities for all in society, such as education \cite{bakerAlgorithmicBiasEducation2022}, healthcare \cite{obermeyerDissectingRacialBias2019}, employment \cite{kochlingDiscriminatedAlgorithmSystematic2020}, justice \cite{malekCriminalCourtsArtificial2022} and commerce \cite{gautierAIAlgorithmsPrice2020}.  In the context of the software engineering discipline, testing software for non-discrimination is a form of non-functional testing \cite{galhotraFairnessTestingTesting2017,brunSoftwareFairness2018}, since biased and unfair decisions have a serious impact on software quality of use, where societal and ethical risks have been recognised in recent standards developments \cite{isoISOIEC250192023,isoISOIEC250592023} and scientific proposals \cite{vetroDataQualityApproach2021}. 
In addition, the introduction of the AI Act includes proof of non-discrimination for high-risk systems \cite{deckImplicationsAIAct2024}.

This study is an audit of an online software system in the Italian car insurance industry: it is ethically and legally imperative to ensure that the software automates and differentiates pricing equitably. We update and extend an original paper by Fabris et al. \cite{fabrisAlgorithmicAuditItalian2021a}, which quantified the impact of gender and birthplace on insurance pricing in an online system. We extend the original study by including additional demographic attributes, which results in a larger and more nuanced test protocol.
This research highlights the importance of testing for non-discrimination in automated pricing algorithms and, more broadly, in software systems that — as the ACM Code of Ethics and Professional Conduct recognises — are “\textit{integrated with everyday activities such as commerce, travel, government, healthcare and education}” \cite{acmcode2018taskforceACMCodeEthics2018} (Article 3.7). 

The remainder of the paper is organized as follows: in Section \ref{sec:rel-work}, we frame the regulations for the domain of interest, and we describe related work in the field; in Section \ref{sec:methodology}, we describe the research questions and experiment design; we report on results in Section \ref{sec:results} and discuss them in Section \ref{sec:discussion}; in Section \ref{sec:threats}, we discuss the potential threats to validity; finally, in Section \ref{sec:conclusions}, we summarize the findings and explore future research directions.

\section{Background and related work}
\label{sec:rel-work}
The Italian car insurance market is regulated by the Italian Insurance Supervisory Authority (IVASS), which collaborates with international organizations like the International Association of Insurance Supervisors (IAIS) and the European Insurance and Occupational Pensions Authority (EIOPA) to ensure market stability and compliance with international standards. This framework prioritizes consumer protection and prohibits discriminatory practices in premium calculations based on personal characteristics. 
As a matter of fact, following a 2011 ruling by the European Court of Justice \cite{europeancourtofjusticeAssociationBelgeConsommateurs2011}, insurers cannot use gender when setting rates. The Article 16 of the European Directive 2021/2118 \cite{counciloftheeuropeanunionCouncilDirective20002000} stipulates that no premium surcharges may be applied on the basis of nationality.

The role of comparison websites is to act as an intermediary between customers and insurance providers. Generally, the insurance companies cover the charges while the customers use the services for no fee. Comparison websites are crucial in the Italian insurance market: during 2022, 54\% of people who took out car or motorbike insurance consulted a comparator\footnote{\href{https://assicurazioni.segugio.it/news-assicurazioni/polizze-auto-e-moto-la-comparazione-si-conferma-lo-strumento-di-riferimento-degli-italiani-00037593.html}{https://assicurazioni.segugio.it/news-assicurazioni/polizze-auto-e-moto-la-comparazione-si-conferma-lo-strumento-di-riferimento-degli-italiani-00037593.html}}. They mediate access to insurance not only by aggregating insurance options, but also by customizing offers by algorithmic optimization. Ongoing monitoring of these websites is essential to ensure they provide unbiased and accurate information, fostering a fair and competitive market.

Algorithmic audits draw from social sciences and can be defined as the collection and analysis of outcomes from an algorithm and a system, to evaluate its accountability \cite{goodmanellenp.ALGORITHMICAUDITINGCHASING2023}. The audit typically simulates a mock user population, with the objective of finding undesired patterns in models of interest \cite{vecchioneAlgorithmicAuditingSocial2021}.
Algorithmic audits have been frequently used to identify potential discrimination by AI systems. A largely influential work in the field of algorithmic audits is that conducted by Raji and Boulamwini, who described Gender Shades, an algorithmic audit of gender and skin type performance disparities in commercial facial analysis models \cite{rajiActionableAuditingInvestigating2019}. The study was replicated after three years by the same authors \cite{rajiActionableAuditingRevisited2022}, highlighting the importance of revisiting the algorithms even long after the original audits have been performed.
Other domains have been interested in algorithmic audits: e.g., Liu et al. have applied auditing to AI applications in the medical domain, to uncover potential algorithmic errors in the context of a clinical task and anticipate their potential consequences \cite{liuMedicalAlgorithmicAudit2022}. 
The auditing process can also involve other techniques than the simulation of inputs. Shen et al. have described the concept of everyday user algorithmic auditing, a process in which users detect, understand, and interrogate problematic machine behaviours via their day-to-day interactions with algorithmic systems \cite{shenEverydayAlgorithmAuditing2021}.
Various institutions have introduced bias auditing systems, as in the case of the New York City Council \cite{thenewyorkcitycouncilLocalLawAmend2021}, reinforcing the need for clearer definitions and metrics \cite{grovesAuditingWorkExploring2024}. In general, the importance of the audit process in exposing the limitations of deployed systems is widely recognised \cite{conitzerTechnicalPerspectiveImpact2022}.

The foundation of this research stems from a 2021 study \cite{fabrisAlgorithmicAuditItalian2021a}, which delves into the Italian car insurance industry. This work uncovers significant insights into the influence of \textit{gender} and \textit{birthplace} on car insurance pricing. Despite regulations prohibiting discriminatory practices, the audit revealed that algorithmic-mediated prices are still influenced by these factors. 
Specifically, foreign-born drivers and individuals from certain Italian cities faced price disadvantages, with Laos drivers being charged up to 1000 € more than drivers with similar profiles in Milan. Additionally, user profiles labelled by the platform as risky received fewer quotes.
Building upon this foundational work, the present article describes the evolution of discriminatory and opaque practices in car insurance pricing.

\section{Methodology}\label{sec:methodology}
This section outlines the research methodology used in this study: firstly we illustrate the research questions, then we report on the experiment design. Finally, we describe the data collection process.

\vspace{-3pt}
\subsection{Goal and research questions}\label{subsec:rq}
The goal of the research is defined with the GQM (Goal-Question-Metric \cite{rinivansolingenGoalQuestionMetric2002}) template as follows: \textbf{test} pricing algorithms \textbf{for the purpose of} auditing an online software system \textbf{with respect to} non-discrimination \textbf{from the point of view} of the users \textbf{in the context of} the Italian car insurance industry.
Two research questions stem from this goal. 

\textbf{RQ1: Do protected attributes (\textit{gender}, \textit{birthplace}, \textit{age}) and socio-demographic attributes (\textit{city}, \textit{marital status}, \textit{education}, \textit{profession}) directly influence quoted premiums?}
This question would determine whether protected attributes\footnote{Protected attributes are selected among the ones identified through Article 21 \textit{"Non-discrimination"} of the EU Charter of Fundamental Rights \cite{europeanunionagencyforfundamentalrightsEUCharterFundamental2015}.} and socio-demographic attributes are related to the insurance quotes provided to the users. This question aims at investigating whether prices vary for similar profiles that differ in only one attribute. 

\textbf{RQ2: Do protected attributes (\textit{gender}, \textit{birthplace}, \textit{age}), socio-\hspace*{0pt}demographic attributes (\textit{city}, \textit{marital status}, \textit{education}, \textit{profession}) and driving attributes (\textit{car}, \textit{km driven}, \textit{class}) influence the number of quotes presented to the user?}
This question focuses on how often car insurance companies appear in search results for different driver profiles. The rationale is to investigate whether profile characteristics expose users to fewer offers, suggesting discrimination when it comes to the availability of quotes among insurance options for them.

\vspace{-3pt}
\subsection{Analysis method}

We performed a preliminary exploratory analysis by analysing the prices' distribution for each attribute value.
As an example, we assessed the spread between the average premium for male and female drivers.
The preliminary exploratory analysis is presented in Section A of the Supplementary materials\footnote{\href{https://anonymous.4open.science/r/RCA-audit-D049/supplementary\_materials.pdf}{https://anonymous.4open.science/r/RCA-audit-D049/supplementary\_materials.pdf}}.

\subsubsection{Discrimination Analysis (RQ1)}\label{subsec:rq1_methodology}
To investigate whether protected attributes directly influence quoted premiums, we analysed the distribution of the price differences $\delta$ for pairs of profiles that differ only in one attribute. The statistical reliability of the median was checked using a sign test with $\mu_0=0$ and considering significant p-values below the $0.05$ threshold. We analysed the top1 and the top5 quotes. 
The top1 analysis concentrates exclusively on the most affordable quote for each profile: this is particularly relevant from the viewpoint of a person who is primarily concerned with the lowest insurance cost.
In contrast, the top5 analysis compares the averages of the five cheapest quotes for every profile, catching a deeper view of the insurance rates for a specific profile.

\begin{table*}[ht]
\footnotesize
\centering
\caption{Features under analysis. Italics highlight the difference from the original study. Abbreviations are shown in brackets. Differences are motivated in Section \ref{subsec:data-collection}.}
\label{tab:variables}
\begin{tabular}{|p{2.2cm}|p{3.5cm}|p{6.3cm}|}
\hline
\textbf{Feature} & \textbf{Values tested in the original study} & \textbf{Values tested in this study} \\ \hline
Gender & Male, Female & Male, Female \\ \hline
Birthplace & Milan, Rome, Naples, \textit{Romania}, \textit{Ghana}, \textit{Laos} & Milan (MI), Rome (RO), Naples (NA), \textit{Morocco (MA)}, \textit{China (CN)} \\ \hline
Age & \textit{18}, 25, 32 & 25, 32 \\ \hline
City & Milan, \textit{Rome}, Naples & Milan (MI), Naples (NA) \\ \hline
Marital Status & \textit{Not present} & \textit{Married (Mar), Single (Sin), Widow (Wid)} \\ \hline
Educational qualification & \textit{Not present} & \textit{Master (MSc), Without a qualification (Waq)} \\ \hline
Profession & \textit{Not present} & \textit{Employee (Emp), Looking for a job (Lfaj)} \\ \hline
Car type & Old, Large Engine, Diesel; New, Small Engine, Petrol & Old, Large Engine, Diesel (OLED); New, Small Engine, Petrol (NSEP) \\ \hline
Km traveled in one year & 10000, 30000 & 10000, 30000 \\ \hline
Class & 1, 4, 9, \textit{14}, 18, \textit{None} & 1, 4, 9, 18 \\ \hline \hline
Total queries & 2160 & 7680 \\ \hline 
\end{tabular}
\end{table*}
\begin{table}[ht]
\footnotesize
\centering
\caption{Number of quotes (\#Q) and frequency of quotes (Freq.) for each company (Comp.) in the collected data.}
\label{tab:companies}
\begin{tabular}{||l|r|r||l|r|r||l|r|r||l|r|r||}
\hline
\textbf{Comp.} & \textbf{\#Q} & \textbf{Freq.} & \textbf{Comp.} & \textbf{\#Q} & \textbf{Freq.} & \textbf{Comp.} & \textbf{\#Q} & \textbf{Freq.} & \textbf{Comp.} & \textbf{\#Q} & \textbf{Freq.} \\ \hline

C1/a & 7.680 & 100\% & C2/a & 2.818 & 36\%  & C3/b &  482   &  6\% & C5/a & 5.853  & 76\%  \\ \hline
C1/b & 1.831 & 24\%  & C2/b & 1.960 & 24\%  & C3/c & 1.844 & 24\%  & C5/b & 1.825  & 24\% \\ \hline
C1/c & 7.660 & 99\%  & C2/c & 177   &  3\%  & C3/d & 1.419  & 19\% &  C6/a & 499   &  6\% \\ \hline
C1/d & 2.792 & 35\%  & C3/a & 1.894 & 25\%  & C4/a & 3.912  & 64\% & & &  \\ \hline

\end{tabular}
\end{table}
%


\vspace{-3pt}
\subsubsection{Output Variability (RQ2)}\label{subsec:rq2_methodology}
To investigate whether specific attributes directly influence the number of quoted premiums offered to users, we observed the proportion of profiles for which each company was present in the offers provided by the comparison site. E.g., to examine the presence of a company for \textit{gender}=male, we calculated the ratio of male profiles for which the company offered a quote to the total number of male profiles.
By computing this ratio, we verify whether an attribute influences the number of quotes presented to the users.

\subsection{Experiment Design and Data Collection}\label{subsec:data-collection}
The data source is a popular italian comparison website.
We collected about 700 insurance policy quotes per day for two weeks, in the second half of January 2024, for a total of 7680 queries. Automatic web scraping with IP rotation with different proxies was integrated by manual data collection.
Three potential sources of variability could be identified: the evolution of insurance company models and pricing schemes, the session duration effects, and the A/B testing conducted by insurance companies and/or by the online software itself.
To ensure the robustness of the data collected, we used a double nested randomization approach \cite{fabrisAlgorithmicAuditItalian2021a}.
The first level of randomization (inner loop) involved randomizing the order in which profiles were queried on the website, so that each profile combination (e.g. birthplace RO and gender M) had an equal chance of being queried at any given time.
The second level of randomization (outer loop) involved randomizing the order of query blocks, as each block consisted of a set of profiles that were to be queried together. 

A control group was included in the study design. We collected control pairs performing two identical queries. This group was expected to not be affected by the variables being tested, and so can be used as a baseline against which the results of the profiles could be compared.
If the price differences for protected pairs are significantly different from those of control pairs in terms of \textit{frequency} (percentage of records with a difference less than $\pm$5 €, i.e. Ties$_5$) and \textit{magnitude} (assessable by observing percentiles), this indicates a bias in the insurance pricing mechanism. Control pairs are also helpful to manage the risk of having price fluctuations related to non-modelled factors (e.g. price updates).

We report a summary of the variables used in Table \ref{tab:variables}, comparing them with the original study. We removed some values analysed in the original paper because, despite IP address rotation, the platform blocked automatic data collection and web tests were integrated with manual tests. We have removed the less relevant values: \textit{Rome} as place of residence; \textit{age} 18, because at that age in Italy you can only drive a car with a low power to weight ratio; \textit{Romania} because it is an EU country; \textit{class} 14 because in the original study it turned out to represent an intermediate value between classes 9 and 18. We have changed the two non-EU countries to MA and CN because people with these nationalities make up the largest African and Asian (respectively) communities in Italy\footnote{\href{https://esploradati.istat.it/databrowser/\#/en/dw/categories/IT1,POP,1.0/POP_FOREIGNIM/DCIS_POPSTRCIT1/IT1,29_317_DF_DCIS_POPSTRCIT1_1,1.0}{istat.it - Italy, regions, provinces - Country of citizenship (Frequency: Annual, Indicator: Foreign census population on 1st January, Time: 2023)}}.

We have released the data collected and the Python scripts used to analyse them in an open-source repository \footnote{\href{https://anonymous.4open.science/r/RCA-audit-D049/}{https://anonymous.4open.science/r/RCA-audit-D049/}}.

\section{Results}
\label{sec:results}
The data collected from the queries include six different companies that were offering up to four insurance products for each query.
The insurance companies are labelled progressively from C1 to C6, with the corresponding services being differentiated as /a, /b, /c, /d. 
Different services represent different commercial offers from the same company (e.g. with or without GPS tracking).
Examination of the compiled insurance quotes demonstrates the divergence of frequencies and number of quotes offered by each insurance company and product set (details reported in Table \ref{tab:companies}). 

\subsection{Discrimination Analysis (RQ1)}\label{subsec:rq1_results}
\begin{table}[tb]
\caption{Discrimination analysis Top1. The first value of each pair is the \textit{test value}, while the second value of each pair is the \textit{baseline}. The difference prices are calculated as test minus baseline. Ties$_{5}$ represents the percentage of protected pairs for which quote difference is within a tolerance threshold of $\pm$5 euros; $\eta.05(\delta)$ and $\eta.95(\delta)$ the 5th and the 95thy percentiles, respectively; $\eta.50(\delta)$ is the median; $m(\delta)$ represents the mean; the p-value tests the null hypothesis that the median difference is zero.}
\label{tab:discrimination-analysis-top1}
\centering
\begin{tabular}{p{1.5cm}p{1.8cm}|rrrrrr}
\toprule
               &             & \multicolumn{6}{c}{Top1} \\ 
Attribute      & Pairs    & Ties$_{5}$  & $\eta.05(\delta)$ & $\eta.50(\delta)$ & $\eta.95(\delta)$ & $m(\delta)$ & $p$ \\\hline
Gender & F vs M & 89\% & -14 € & 0 € & 6 € & 4 € & 0.72 \\
Birthplace & RO vs MI & 25\% & 0 € & 8 € & 10 € & 7 € & \textbf{<0.05} \\
Birthplace & NA vs MI & 10\% & -8 € & 29 € & 538 € & 92 € & \textbf{<0.05} \\
Birthplace & MA vs MI & 0\% & -255 € & 125 € & 539 € & 148 € & \textbf{<0.05} \\
Birthplace & CN vs MI & 10\% & -104 € & 103 € & 395 € & 118 € & \textbf{<0.05} \\
Age & 25 vs 32 & 1\% & -1 € & 64 € & 549 € & 141 € & \textbf{<0.05} \\
City & NA vs MI & 31\% & -255 € & 147 € & 1752 € & 367 € & \textbf{<0.05} \\
Mar. Stat. & Sin vs Mar & 79\% & -33 € & 0 € & 83 € & -1 € & \textbf{<0.05} \\
Mar. Stat. & Wid vs Mar & 79\% & -30 € & 0 € & 183 € & 9 € & \textbf{<0.05} \\
Education & WaQ vs MSc & 77\% & -9 € & 0 € & 491 € & 60 € & \textbf{<0.05} \\
Profession & LfaJ vs Emp & 70\% & -377 € & 0 € & 283 € & 24 € & \textbf{<0.05} \\
\multicolumn{2}{l|}{Control pairs (noise)}  & 98\% & 0 € & 0 € & 0 € & 7 € & 1.00 \\ \hline
\end{tabular}
\end{table}

\begin{table}[tb]
\caption{Discrimination analysis on the averaged five cheapest values (columns descriptions in Table \ref{tab:discrimination-analysis-top1}).}
\label{tab:discrimination-analysis-top5}
\centering
\begin{tabular}{p{1.5cm}p{1.8cm}|rrrrrr}
\toprule
               &             & \multicolumn{6}{c}{Top5} \\ 
Attribute      & Pairs    & Ties$_{5}$  & $\eta.05(\delta)$ & $\eta.50(\delta)$ & $\eta.95(\delta)$ & $m(\delta)$ & $p$ \\\hline
Gender      & F vs M & 78\% & -61 € & 0 € & 128 € & 4 € & \textbf{<0.05} \\
Birthplace  & RO vs MI & 71\% & 0 € & 2 € & 9 € & 4 € & \textbf{<0.05} \\
Birthplace  & NA vs MI & 4\% & -199 € & 113 € & 382 € & 128 € & \textbf{<0.05} \\
Birthplace  & MA vs MI & 0\% & -178 € & 252 € & 1187 € & 371 € & \textbf{<0.05} \\
Birthplace  & CN vs MI & 4\% & -75 € & 128 € & 712 € & 200 € & \textbf{<0.05} \\
Age         & 25 vs 32 & 1\% & 0 € & 211 € & 877 € & 285 € & \textbf{<0.05} \\
City        & NA vs MI & 23\% & -346 € & 278 € & 1954 € & 657 € & \textbf{<0.05} \\
Mar. Stat.  & Sin vs Mar & 26\% & -116 € & 10 € & 238 € & 42 € & \textbf{<0.05} \\
Mar. Stat.  & Wid vs Mar & 25\% & -56 € & 35 € & 515 € & 110 € & \textbf{<0.05} \\
Education   & WaQ vs MSc & 29\% & -7 € & 99 € & 896 € & 236 € & \textbf{<0.05} \\
Profession  & LfaJ vs Emp & 24\% & -188 € & 22 € & 769 € & 135 € & \textbf{<0.05} \\
\multicolumn{2}{l|}{Control pairs (noise)}  & 98\% & 0 € & 0 € & 0 € & 10 € & 1.00 \\ \hline
\end{tabular}
\end{table}

Tables \ref{tab:discrimination-analysis-top1} and \ref{tab:discrimination-analysis-top5} presents the results of the discrimination analysis. Negative values indicate that profiles with the test value obtained lower prices; positive values indicate that they obtained higher quotes compared to those obtained by the baseline. We observe that:

\begin{itemize}
    \item \textit{Gender}, confronting females and males profiles, on one hand shows a median equal to zero in both top1 and top5. On the other hand, the difference between $\eta.95(\delta)$ and $\eta.05(\delta)$ raises from 20 € in top1 to 189 € in top5 (0 € in control pairs), while $Ties_5$ drop from 89\% to 78\% (98\% in control pairs).
    \item \textit{Birthplace}, is used to the advantage of drivers born in Milan. Drivers born outside Italy were the most highly rated: on average, drivers born in Morocco are charged 200 € more than otherwise identical drivers born in Milan.
    \item \textit{Age} shows a median difference of 64 € in the analysis of the cheapest offer, this value increases to 211 € when looking at the average of the five cheaper offers.
    \item The \textit{city} attribute reveals better prices in favour of Milan residents. The median and mean values show greater discrimination between Naples and Milan for the City attribute than for the \textit{birthplace} attribute.
    \item \textit{Marital status}: The top1 values show a median and an average equal or close to zero for both pairs, while the 95th percentile shows values 3 to 6 times higher than the 5th percentile. In the top5 analysis, the median and the average are higher than zero, especially for the pair 'widowed vs. married' (35 € and 110 € respectively).
    \item \textit{Education} shows a relevant imbalance in the top5 analysis for the damage to the unqualified (WaQ), with a price difference that is 99 € higher on the median and 236 € higher on the average. For the top1 values this effect is reduced: the median is 0 €, while the average is 60 €.
    \item For \textit{Profession}, analysing the top5 values, jobseekers (Lfaj) received offers that were on median 22 € higher and on average 135 € higher.
\end{itemize}

Taking into account the significant disparities directly caused by protected attributes, we observe that the magnitude of differences in the top5 results, measured as $\eta.95(\delta)-\eta.05(\delta)$, vary from 9€ (birthplace Rome vs Milan) to 2300€ (city Naples vs Milan), compared to a value of 0 € for control pairs. The frequency of Ties5 for top5 is below 5\% for age and all the birthplace pairs except for Rome vs Milan, compared against a value of 98\% for control pairs. Similar, albeit slightly less pronounced, patterns are observed in the top1 results.

\begin{tcolorbox}
\textbf{RQ1}: Protected attributes (\textit{gender}, \textit{birthplace}, \textit{age}) and socio-demographic attributes (\textit{city}, \textit{marital status}, \textit{education}, \textit{profession}) have a direct effect on premiums quoted. \textit{Birthplace}, \textit{age}, \textit{city} and \textit{education} systematically reveal a direction of such influence.
\end{tcolorbox}

\subsection{Output Variability (RQ2)}\label{subsec:rq2_results}
\begin{figure*}[t]
    \centering
    \includegraphics[width=1\linewidth]{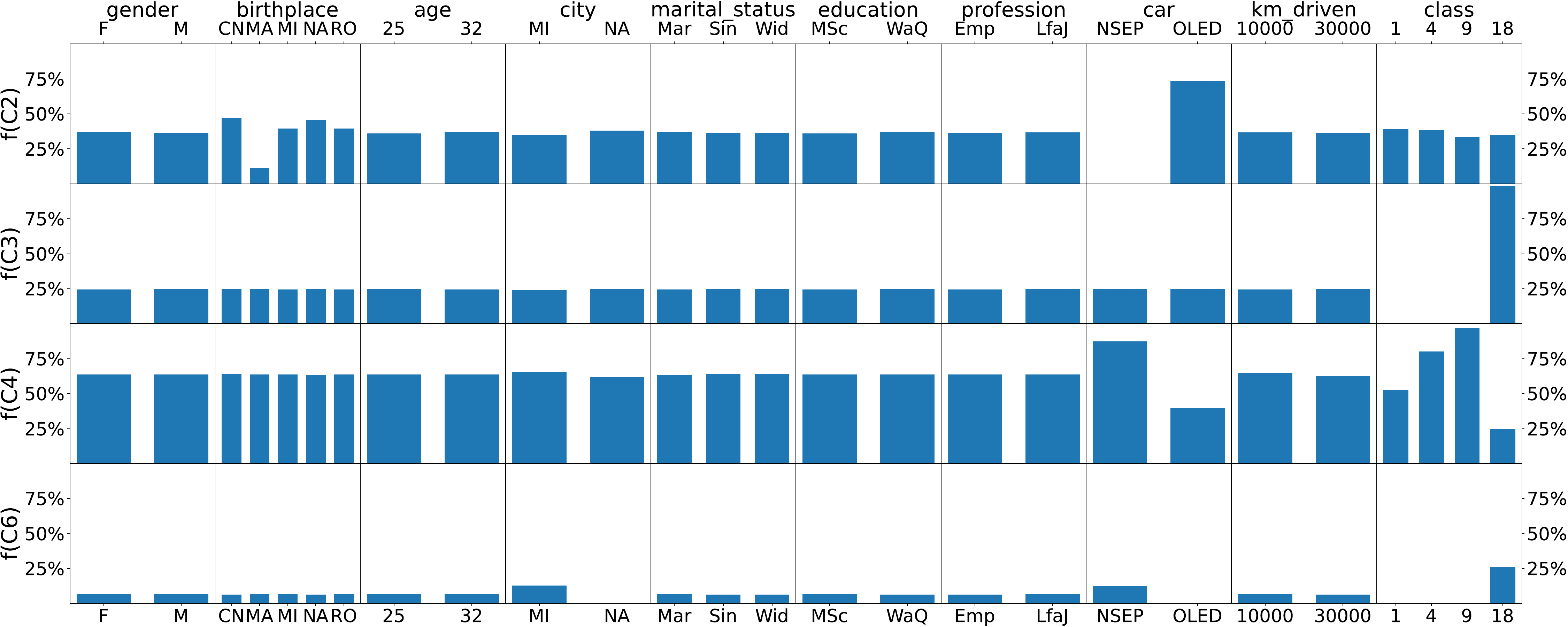}
    \caption{Presence of companies as a percentage of offers for each attribute.}
    \label{fig:output-variability}
\end{figure*}

Figure \ref{fig:output-variability} provides an overview of the percentage of quotes presented by each company, for each value of the attributes. 
Companies \textit{C1} and \textit{C5} are not represented, as they submitted at least one offer for all profiles, i.e. they reached 100\% for each class of all attributes. Regarding the other companies: \textit{C2} appeared for about a third of the profiles, \textit{C3} for about a quarter of the profiles, \textit{C6} offered an insurance only for a small subset of records and \textit{C4} was present in about 60\% of the queries. It is possible to see a few relevant differences for different profiles, but these differences are company-specific with no general trends:
\begin{itemize}
    \item The company \textit{C2} submitted fewer quotes for profiles with \textit{birthplace} setted to Morocco. At the same time, it submitted quotes for about 75\% of profiles with the car type Old, Large Engine, Diesel (OLED), while no quotes appeared for New, Small Engine, Petrol (NSEP). 
    \item The company \textit{C3} was rather balanced in all characteristics, except for risk \textit{class}: it was present in almost all profiles with the highest risk class (18), but was absent for the others.
    \item The company \textit{C4} presents significantly more quotes for NSEP cars than for OLED cars. It also presents a non-uniform distribution of quotes per risk class.
    \item  The company \textit{C6} presents fewer quotes, so it is more difficult to draw relevant conclusions. However, it is possible to notice that it only presented quotes for  the riskier driving class and for profiles living in Milan, and that there are no quotes for OLED cars. 
\end{itemize}

\begin{tcolorbox}
\textbf{RQ2}: \textit{Birthplace} and \textit{city} are factors that influence the number of quotes in two companies. Driver characteristics such as \textit{car} type and \textit{risk class} also expose users to a different number of quotes in four companies.
\end{tcolorbox}

\section{Discussion}
\label{sec:discussion}

\textbf{RQ1: Do protected attributes (\textit{gender}, \textit{birthplace}, \textit{age}) and socio-demographic attributes (\textit{city}, \textit{marital status}, \textit{education}, \textit{profession}) directly influence quoted premiums?}
The original study looked at price discrimination based on \textit{birthplace} and \textit{gender}.
With regard to the first one, we confirm that the \textit{birthplace} has a direct impact on the premium offered: people born in Naples systematically receive higher prices.
This type of discrimination is even more pronounced when we look at the profiles with an Italian birthplace compared to the profiles with a foreign birthplace (in our case, Moroccan and Chinese).
With regard to \textit{gender}, as in the original study, there is no systematic advantage in one direction (male or female).
\textit{Age} is used to systematically discriminate against younger people: this is expected as – on average – the youngest have a higher risk class, however, this is not systematic as the risk class in Italy can be inherited from a family member. Further investigations looking at the intersection of age and risk class are needed.
\textit{City} of residence is the attribute that reveals the greater systematic discrimination: residents in Milan pay, on average, 128 € less than residents in Naples. 
On the one hand, it seems that insurance companies consider driving in Naples to be riskier than driving in Milan. On the other hand, the Italian National Statistics Institute, in its latest report on road accidents (2022), showed that the absolute number of road accidents and the fatality rate per 100,000 inhabitants were higher in Milan than in Naples \cite{istitutonazionaledistatisticaIncidentiStradaliItalia2022}.
\textit{Marital status} and \textit{profession} show a slight discrimination against married, widowed and unemployed profiles. In terms of \textit{education}, those with no qualifications are paying way much more than those with a Master's degree (in the top5 analysis). We can hypothesise that these attributes play the role of a proxy variable for the age attribute, since younger people are considered to be more at risk of car accidents \cite{gomes-francoExplainingAssociationDriver2020}. However, further multivariate distribution analysis should be performed to confirm this hypothesis. 
Nevertheless, the magnitude of the differences confirms that all the protected characteristics analysed play a relevant role in the pricing algorithms.

\textbf{RQ2: Do protected attributes (\textit{gender}, \textit{birthplace}, \textit{age}), socio-\hspace*{0pt}demographic attributes (\textit{city}, \textit{marital status}, \textit{education}, \textit{profession}) and driving attributes (\textit{car}, \textit{km driven}, \textit{class}) influence the number of quotes presented to the user?}
The original study showed that younger people, residents of Naples and drivers with the worst risk class received fewer offers.
With regard to \textit{age}, we do not observe any significant differences between 25 and 32, but it should be noted that we did not test the age of 18 (as described in Section \ref{subsec:data-collection}), which seemed to be the least desirable for insurance companies.
Observing the influence of the city of residence, only one company confirms the pattern of imbalance between Milan and Naples in terms of the number of offers.
As for the claims' history (\textit{class}), the results are controversial.
On the one hand, company \textit{C4} confirms the pattern shown in the original study, i.e. the fact that the riskier profiles receive fewer quotes. 
On the other hand, companies \textit{C3} and \textit{C6} appear only for the highest risk class.
We can hypothesise that some companies are attracted by the possibility of obtaining higher revenues. 

\section{Threats to Validity}
\label{sec:threats}

In this section, we report the main threats to the validity of the study according to classifications available in the literature \cite{feldtValidityThreatsEmpirical2010}. The main \emph{external validity threat} of our study is that the examination has been conducted on only one comparator website. Future studies shall expand the scope to include multiple systems to provide a more comprehensive understanding of the issues. It is not ensured as well whether our results can be applied to other countries or other domains. As \emph{internal validity threats} are concerned, the major ones lie in the selection of the variables that were analysed. Additional variables among those collected by the algorithms could be in fact investigated, including further proxies of socio-economic conditions and intersectional attributes. 

\section{Conclusions and future work}
\label{sec:conclusions}
This study is an extended and updated audit of pricing algorithms used in an online  system comparing car insurance prices in Italy. The analysis confirms and extends the findings of the original study: several demographic variables had a significant impact on pricing, with place of birth emerging as a discriminatory factor, especially for those not born in Italian cities. In terms of non-equal opportunities, driver profiles (e.g., car type) determine the options available to users, and for two companies this was based on protected attributes. 

These results not only demonstrate the importance of verifying fairness in algorithmic pricing mechanisms and, more generally, of continuously testing software services for non-discrimination, but they also show that empirical methods – especially in relation to experiment design in conjunction with testing techniques – are well suited to such goals. As a result, our work is an example of the potential role that empirical software engineering can play in the emerging field of testing algorithms for non-discrimination. 
With this vision in mind, future work in this domain will be devoted to increasing the level of test automation, switching features – including intersectional attributes– and tracking prices over time. 
More generally, we hope that this vision lead to future experiments in the other areas identified by Art. 3.7 of the ACM Code of Ethics and Professional Conduct, where non-discrimination is paramount.

\begin{credits}
\subsubsection{\ackname} This study was carried out within the FAIR - Future Artificial Intelligence Research and received funding from the European Union Next-GenerationEU (PIANO NAZIONALE DI RIPRESA E RESILIENZA (PNRR) – MISSIONE 4 COMPONENTE 2, INVESTIMENTO 1.3 – D.D. 1555 11/10/2022, PE00000013). This manuscript reflects only the authors’ views and opinions, neither the European Union nor the European Commission can be considered responsible for them.
\end{credits}
%
%

\bibliographystyle{splncs04}
\bibliography{06_rca.bib}

\end{document}